\documentclass[aps,prb,preprint,showpacs,showkeys,floatfix]{revtex4}

\usepackage{graphicx,color}
\newcommand{\jwj}[1]{\textcolor{red}{#1}}
\graphicspath{{figs/}}
\begin{document}
\title{The Buckling of Single-Layer MoS$_{2}$ Under Uniaxial Compression}
\author{Jin-Wu Jiang}
    \altaffiliation{Email address: jwjiang5918@hotmail.com}
    \affiliation{Shanghai Institute of Applied Mathematics and Mechanics, Shanghai Key Laboratory of Mechanics in Energy Engineering, Shanghai University, Shanghai 200072, People's Republic of China}

%\date{22 December 2009}
\date{\today}
\begin{abstract}
Molecular dynamics simulations are performed to investigate the buckling of single-layer MoS$_{2}$ under uniaxial compression. The strain rate is found to play an important role on the critical buckling strain, where higher strain rate leads to larger critical strain. The critical strain is almost temperature-independent for $T<50$~K, and it increases with increasing temperature for $T>50$~K owning to the thermal vibration assisted healing mechanism on the buckling deformation. The length-dependence of the critical strain from our simulations is in good agreement with the prediction of the Euler buckling theory.
\end{abstract}

\maketitle
\pagebreak

\section{Introduction}

Molybdenum Disulphide (MoS$_{2}$) is a semiconductor with a bulk bandgap above 1.2~{eV},\cite{KamKK} which can be further manipulated by changing its thickness,\cite{MakKF} or through application of mechanical strain.\cite{FengJ2012npho,LuP2012pccp,ConleyHJ,CheiwchanchamnangijT2013prb} This finite bandgap is a key reason for the excitement surrounding MoS$_{2}$ as compared to graphene as graphene has a zero bandgap.\cite{NovoselovKS2005nat,CastroNAH,PereiraVM}  Because of its direct bandgap and also its well-known properties as a lubricant, MoS$_{2}$ has attracted considerable attention in recent years.\cite{WangQH2012nn,ChhowallaM} For example, Radisavljevic et al.\cite{RadisavljevicB2011nn} demonstrated the application of single-layer MoS$_{2}$ (SLMoS$_{2}$) as a good transistor. The strain and the electronic noise effects were found to be important for the SLMoS$_{2}$ transistor.\cite{ConleyHJ,SangwanVK,Ghorbani-AslM,CheiwchanchamnangijT}

Besides the electronic properties, several recent works have addressed the thermal and mechanical properties of SLMoS$_{2}$\cite{gomezAM2012,Castellanos-GomezA2012nrl,HuangW,BertolazziS,CooperRC2013prb1,CooperRC2013prb2,JiangJW2013mos2,VarshneyV,LiuX2013apl,Castellanos-GomezA2013nl} Recently, we have parametrized the Stillinger-Weber potential for SLMoS$_{2}$.\cite{JiangJW2013sw} Based on this Stillinger-Weber potential, we derived an analytic formula for the elastic bending modulus of the SLMoS$_{2}$, where the importance of the finite thickness effect was revealed.\cite{JiangJW2013bend} We have also shown that the MoS$_{2}$ resonator has much higher quality factor than the graphene resonator.\cite{JiangJW2013mos2resonator}

As an important mechanical phenomenon, the buckling of graphene has attracted lots of attention in past few years.\cite{LuQ2009ijam,PatrickWJ2010jctn,SakhaeePA2009cms,PradhanSC2009cms,PradhanSC2009plsa,FrankO2010acsnn,FarajpourA2011pe,TozziniV2011jpcc,RouhiS2012pe,GiannopoulosGI2012cms,Neek-AmalM2012apl,ShenH2013apl} Compared to graphene, the bending modulus of SLMoS$_{2}$ is higher by a factor of seven due to its finite thickness,\cite{JiangJW2013bend} yet the in-plane bending stiffness in SLMoS$_{2}$ is smaller than graphene by a factor of five.\cite{JiangJW2013sw} As a result, the SLMoS$_{2}$ should be more difficult to be buckled than graphene, according to the Euler buckling theory, which says that the buckling critical strain is proportional to the bending modulus to in-plane stiffness ratio.\cite{TimoshenkoS1987} This advantage would benefit for the application of SLMoS$_{2}$ in some mechanical devices, where the avoidance of buckling is desirable. However, the study of the buckling for the SLMoS$_{2}$ is still lacking, and is thus the focus of the present work.

In this paper, we study the buckling of the SLMoS$_{2}$ under uniaxial compression via molecular dynamics (MD) simulations. We find that the critical buckling strain increases linearly with the increase of the strain rate, and this strain rate effect becomes more important for longer SLMoS$_{2}$. The critical strain is almost temperature-independent in low-temperature regions, but it increases considerably with increasing temperature for temperatures above 50~K, because the buckling deformation is repaired by the thermal vibration at higher temperatures. We show that the critical strain depends on the length ($L$) as $\epsilon_{c}\propto 40.6/L^{2}$, which is in good agreement with the prediction of $\epsilon_{c}\propto 43.52/L^{2}$ from the Euler buckling theory.

\section{Euler Buckling Theory}
\begin{figure}[htpb]
  \begin{center}
    \scalebox{1}[1]{\includegraphics[width=8cm]{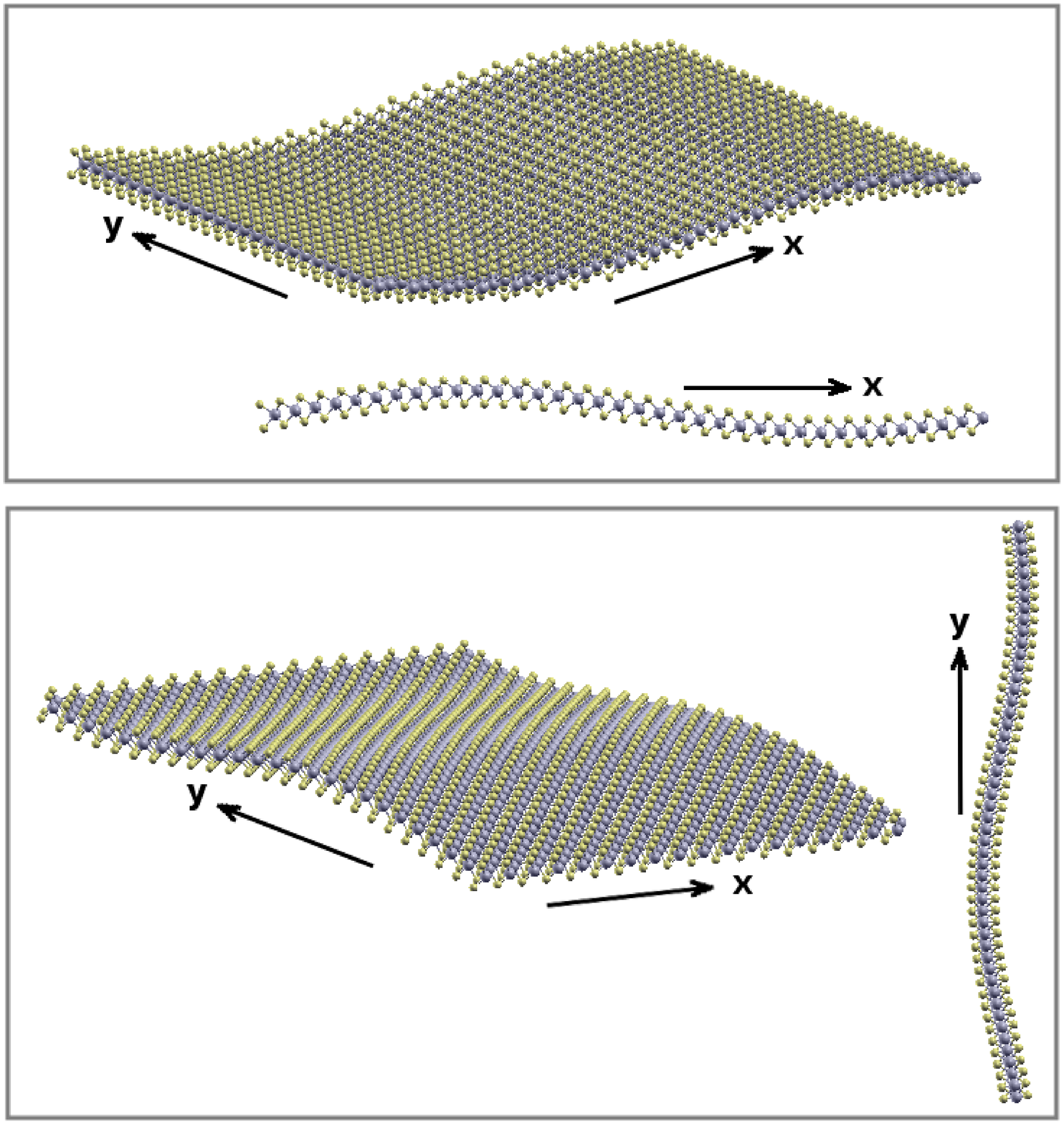}}
  \end{center}
  \caption{(Color online) The two lowest-frequency phonon normal modes in the SLMoS$_{2}$ of dimension $100\times 80$~{\AA}, with periodic boundary condition in both $x$ and $y$ directions. The position of atom $i$ plotted in the figure is $\vec{r}_{i}=\vec{R}_{i}+C\vec{u}_{i}$, where $\vec{R}_{i}$ is the equilibrium position, $\vec{u}_{i}$ is the component of atom $i$ in the eigen vector of the normal mode. $C$ is a scaling factor for the convenience of illustration. The eigen vector is $\vec{u}_{i}=\vec{e}_{z}\sin \frac{2\pi x}{L_{x}}$ for the first lowest-frequency mode in the top panel, and $\vec{u}_{i}=\vec{e}_{z}\sin \frac{2\pi y}{L_{y}}$ for the second lowest-frequency mode in the bottom panel.}
  \label{fig_buckling_mode}
\end{figure}

We first present some basic knowledge for the Euler buckling theory. Similar as the visual displacement method used in Ref.~\onlinecite{TimoshenkoS1987}, let's assume that the buckling happens by exciting one normal (phonon) mode $\left(m,n\right)$ in SLMoS$_{2}$; i.e. the system is deformed into the shape corresponding to one of its normal mode,
\begin{eqnarray}
w & = & w_{mn}=a_{mn}\sin\frac{m\pi x}{L_{x}}\sin\frac{n\pi y}{L_{y}},
\label{eq_eigvec}
\end{eqnarray}
where $a_{mn}$ is the amplitude. $L_{x}$ and $L_{y}$ are the length in $x$ and $y$ directions. Integers $m$, $n=$0, 2, 4, ... for periodic boundary condition. Because the buckling mode is in close relation with the normal mode, we show the first two lowest-frequency normal modes with integers $(m,n)=$ (2,0) and (0,2) in the top and bottom panels in Fig.~\ref{fig_buckling_mode}. The periodic boundary condition is applied in $x$ and $y$ directions. The eigen vector is obtained by diagonalizing the dynamical matrix $K_{ij}=\partial^{2}V/\partial x_{i}\partial x_{j}$, which is obtained numerically by calculating the energy change after a small displacement of the $i$-th and $j$-th degrees of freedom.

After some simple algebra, we get the general strain energy corresponding to a general mode $(m,n)$ as
\begin{eqnarray}
V_{S} & = & \frac{1}{2}\int\int\left[N_{x}\left(\frac{\partial w}{\partial x}\right)^{2}+N_{y}\left(\frac{\partial w}{\partial y}\right)^{2}+2N_{xy}\frac{\partial w}{\partial x}\frac{\partial w}{\partial y}\right]dxdy\nonumber\\
 & = & \frac{\pi^{2}}{8}L_{x}L_{y}N_{x}m^{2}\frac{a_{mn}^{2}}{L_{x}^{2}},
\label{eq_VS}
\end{eqnarray}
where $N_{x}$ is the force in the $x$ direction. The bending energy corresponding to the mode $(m,n)$ is
\begin{eqnarray}
V_{D} & = & \frac{D}{2}\int\int\left\{ \left(\frac{\partial^{2}w}{\partial x^{2}}+\frac{\partial^{2}w}{\partial y^{2}}\right)^{2}-2\left(1-\nu\right)\left[\frac{\partial^{2}w}{\partial x^{2}}\frac{\partial^{2}w}{\partial y^{2}}-\left(\frac{\partial^{2}w}{\partial x\partial y}\right)^{2}\right]\right\} dxdy\nonumber\\
 & = & \frac{\pi^{4}L_{x}L_{y}}{8}Da_{mn}^{2}\left[\left(\frac{m}{L_{x}}\right)^{2}+\left(\frac{n}{L_{y}}\right)^{2}\right]^{2},
\label{eq_VB}
\end{eqnarray}
where $\nu$ is the Poisson ratio and $D$ is the bending modulus.

Within the configuration just before buckling, the strain energy is maximum and there is no bending energy. After buckling, this maximum strain energy is fully converted into the bending energy; i.e. $V_{S}=V_{B}$. From this equation, we get the critical force for buckling
\begin{eqnarray*}
N_{x} & = & -\frac{\pi^{2}D}{L_{y}^{2}}\left(\frac{mL_{y}}{L_{x}}+\frac{n^{2}L_{x}}{mL_{y}}\right)^{2}.
\end{eqnarray*}
Recall that $N_{x}=C_{11}\epsilon$ with $C_{11}$ as the in-plane stiffness, we get the critical strain for buckling,
\begin{eqnarray}
\epsilon_{c} & = & -\frac{\pi^{2}D}{C_{11}L_{y}^{2}}\left(\frac{mL_{y}}{L_{x}}+\frac{n^{2}L_{x}}{mL_{y}}\right)^{2}.
\label{eq_strain1}
\end{eqnarray}
Obviously, the minimum value of $\epsilon_{c}$ is chosen at $(m,n)=(2,0)$; i.e. the buckling happens by deforming the SLMoS$_{2}$ into the shape of the first lowest-frequency normal mode shown in the top panel of Fig.~\ref{fig_buckling_mode}. We note that the choose of $m=2$ instead of $m=1$ here is the result of the constraint from the periodic boundary condition applied in the $x$ direction. The Stillinger-Weber potential gives a bending modulus\cite{JiangJW2013bend}  $D=9.61$~{eV} and the in-plane tension stiffness\cite{JiangJW2013sw} $C_{11}=139.5$~{Nm$^{-1}$} for the SLMoS$_{2}$. Using these two quantities, we get an explicit formula for the critical strain of SLMoS$_{2}$,
\begin{eqnarray}
\epsilon_{c}  =  -\frac{4\pi^{2}D}{C_{11}L_{x}^{2}} =  -\frac{43.52}{L_{x}^{2}} \equiv -\frac{43.52}{L^{2}}.
\label{eq_strain2}
\end{eqnarray}
Hereafter, we will use $L$ to denote the length of the SLMoS$_{2}$ instead of $L_{x}$.

\section{Results and Discussions}
After the representation of the Euler buckling theory, we are now performing MD simulations to study the buckling of SLMoS$_{2}$. All MD simulations in this work are performed using the publicly available simulation code LAMMPS~\cite{PlimptonSJ,Lammps}, while the OVITO package was used for visualization in this section~\cite{ovito}. The standard Newton equations of motion are integrated in time using the velocity Verlet algorithm with a time step of 1~{fs}. \jwj{The interaction within MoS$_{2}$ is described by the Stillinger-Weber potential, where the parameters for this potential have been fitted to the phonon dispersion of single-layer and bulk MoS$_{2}$.\cite{JiangJW2013sw} The phonon dispersion is closely related to some mechanical quantities like Young's modulus and some thermal properties, so this parameter set can give a good description for the mechanical and thermal properties of the single-layer MoS$_{2}$.} Periodic boundary conditions are applied in the two in-plane directions, and the free boundary condition is applied in the out-of-plane direction.  \jwj{Our simulations are performed as follows.  First, the Nos\'e-Hoover\cite{Nose,Hoover} thermostat is applied to thermalize the system to a constant pressure of 0, and a constant temperature of 1.0~{K} within the NPT (i.e. the particles number N, the pressure P and the temperature T of the system are constant) ensemble, which is run for 100~ps. The SLMoS$_{2}$ is then compressed in the x-direction within the NPT ensemble, which is also maintained through the Nos\'e-Hoover thermostat.} \jwj{The SLMoS$_{2}$ is compressed unaxially along the $x$ direction by uniformly deforming the simulation box in this direction, while it is allowed to be fully relaxed in lateral directions during the compression.} 

\begin{figure}[htpb]
  \begin{center}
    \scalebox{1}[1]{\includegraphics[width=8cm]{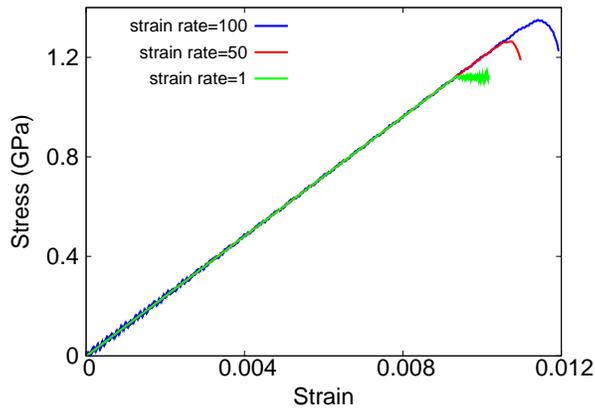}}
  \end{center}
  \caption{(Color online) Stress strain relation under compression for SLMoS$_{2}$ with length 60~{\AA} and width 40~{\AA} at 1.0~{K}. The curve drops sharply at a critical strain $\epsilon_{c}=0.00939$ when the SLMoS$_{2}$ is compressed at a strain rate $\dot{\epsilon}=1.0$~{$\mu$s$^{-1}$}.}
  \label{fig_stress_strain}
\end{figure}

We shall examine the strain rate effect on the critical strain. Fig.~\ref{fig_stress_strain} shows the stress strain relation under compression for SLMoS$_{2}$ with length 60~{\AA} and width 40~{\AA} at 1.0~{K} low temperature. The buckling phenomenon happens when the SLMoS$_{2}$ is compressed with the critical strain $\epsilon_{c}$, which corresponds to a sharp drop in the curve. The critical strain $\epsilon_{c}=0.00939$, 0.01073, and 0.01141 correspond to the strain rate of $\dot{\epsilon}=1.0$~{$\mu$s$^{-1}$}, 10.0~{$\mu$s$^{-1}$}, and 100.0~{$\mu$s$^{-1}$}. The critical strain increases with increasing strain rate. It is because the system is compressed so fast that the relaxation time is not long enough for the occurrence of the buckling phenomenon. 

\begin{figure}[htpb]
  \begin{center}
    \scalebox{1}[1]{\includegraphics[width=8cm]{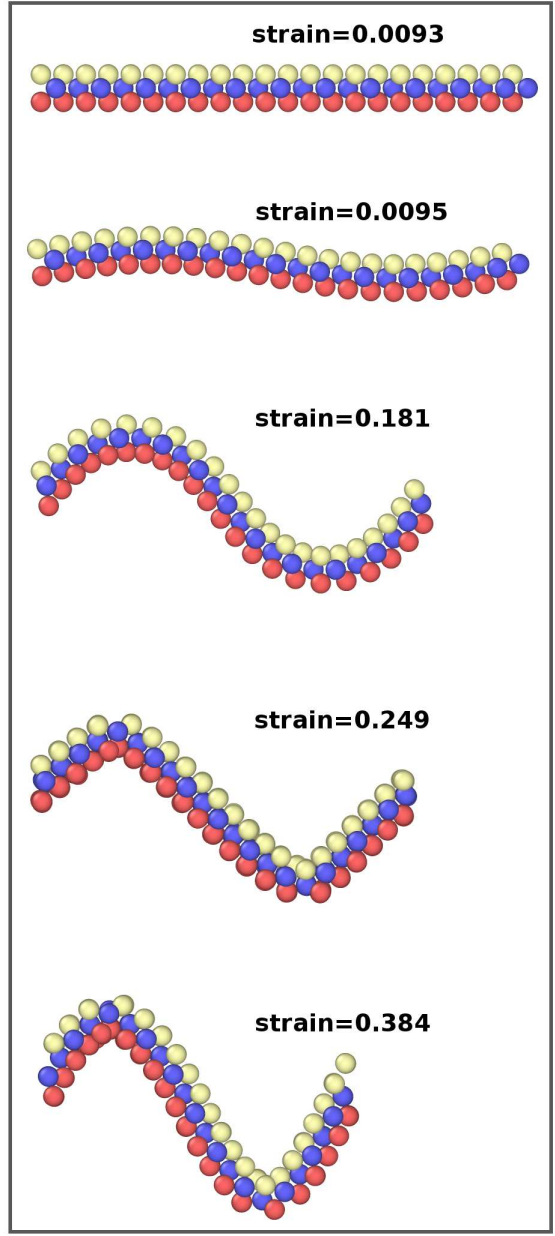}}
  \end{center}
  \caption{(Color online) The structure evolution for SLMoS$_{2}$ with length 60~{\AA} and width 40~{\AA} at 1.0~{K}. The SLMoS$_{2}$ is compressed at a strain rate $\dot{\epsilon}=1.0$~{$\mu$s$^{-1}$}. The buckling happens at $\epsilon=0.00939$. The structure transforms from sinuous into zigzag at $\epsilon=0.249$.}
  \label{fig_cfg_buckling}
\end{figure}

\jwj{Fig.~\ref{fig_cfg_buckling} shows the configuration evolution for the SLMoS$_{2}$. The SLMoS$_{2}$ is compressed at a strain rate $\dot{\epsilon}=1.0$~{$\mu$s$^{-1}$}. The buckling happens at $\epsilon=0.00939$. The first normal mode buckling is observed at the critical strain, as this mode has the lowest exciting energy. The figure shows that MoS$_{2}$ is further deformed with more strain applied. However, the deformation always follows the shape of the first normal mode, and no high-order normal mode is observed. It indicates that the strain energy stored in the buckled SLMoS$_{2}$ (with first normal mode buckling) is not large enough to jump from the first normal mode to high-order normal mode. It is because this jumping requires the reverse of a ripple (with large bending curvature) in the first normal mode, which is forbidden by high potential threshold. It is quite interesting that the structure transforms from sinuous into zigzag at $\epsilon=0.249$, due to stress concentration at the two valleys of the sinuous shape.}

\begin{figure}[htpb]
  \begin{center}
    \scalebox{1}[1]{\includegraphics[width=8cm]{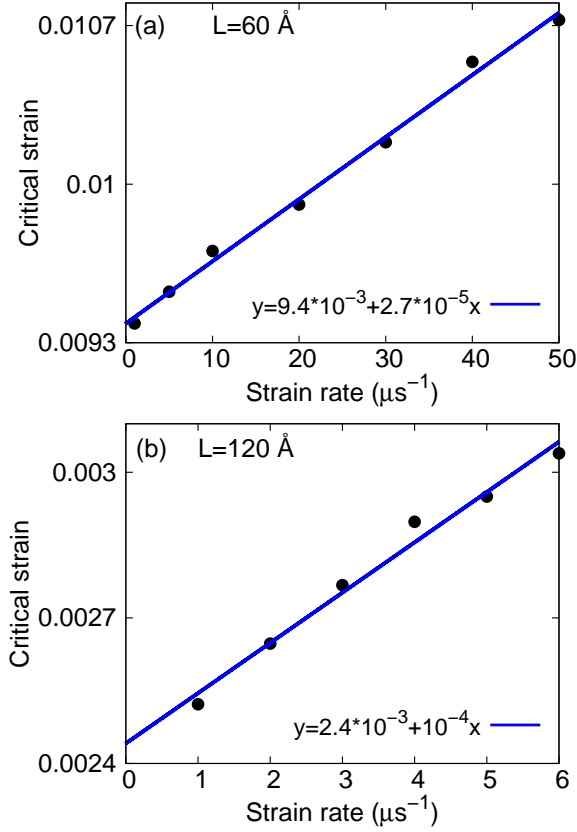}}
  \end{center}
  \caption{(Color online) The strain rate effect on the critical strain for SLMoS$_{2}$ with width 40~{\AA} at 1.0~K. The critical strain increases linearly with increasing strain rate for SLMoS$_{2}$ with length $L=60$~{\AA} in (a) and $L=120$~{\AA} in (b). Note that the slope ($2.7\times 10^{-5}$~{$\mu$s}) of the fitting line in the shorter system in (a) is only a quarter of the slope of $1.0\times 10^{-4}$~{$\mu$s} in the longer MoS$_{2}$ in (b), which implies that the strain rate has stronger effect for longer MoS$_{2}$.}
  \label{fig_srate}
\end{figure}

To further examine the strain rate effect on the critical strain, we perform systematic simulations for the buckling of the SLMoS$_{2}$ under compression with different strain rate. Fig.~\ref{fig_srate} shows the critical strain for SLMoS$_{2}$ with width 40~{\AA} at 1.0~K. The length of the SLMoS$_{2}$ is 60~{\AA} in panel (a) and 120~{\AA} in panel (b). In both systems, the critical strain increases linearly with increasing strain rate. The simulation data are fitted to a linear function $y=a+bx$, where the coefficient $a=9.4\times 10^{-3}$ and $2.4\times 10^{-3}$ can be regarded as the exact value (with $\dot{\epsilon}\rightarrow 0.0$) for the critical strain in these two SLMoS$_{2}$. Using a strain rate of 1.0~{$\mu$s$^{-1}$}, we get the critical strains 0.00938 and 0.00252 for SLMoS$_{2}$ with $L=60$~{\AA} and 120~{\AA}. As a result, the error due to using a finite strain rate of 1.0~{$\mu$s$^{-1}$} is 0.2\% and 5\% for these two SLMoS$_{2}$. For the coefficient $b$, this slope of the fitting line in the shorter system in panel (a) is $2.7\times 10^{-5}$~{$\mu$s}, which is only a quarter of the slope of $1.0\times 10^{-4}$~{$\mu$s} in the longer MoS$_{2}$ in panel (b). It indicates that the strain rate has stronger effect for longer SLMoS$_{2}$, because the frequency of the buckling mode (first lowest-frequency normal mode) is lower in longer system, leading to longer response time. That is the buckling mode in a longer SLMoS$_{2}$ requires longer relaxation time for its occurrence.

\begin{figure}[htpb]
  \begin{center}
    \scalebox{1}[1]{\includegraphics[width=8cm]{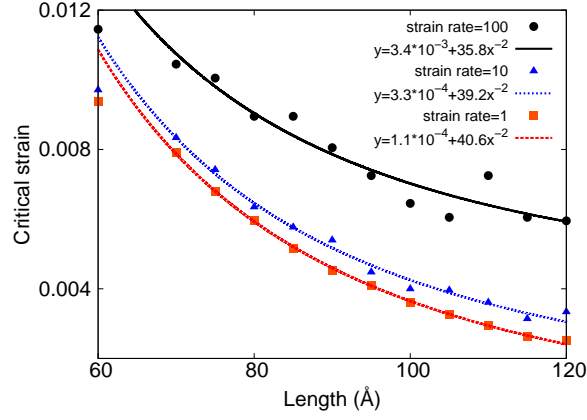}}
  \end{center}
  \caption{(Color online) The length dependence of the critical strain in SLMoS$_{2}$ with width 40~{\AA} at 1.0~K. The system is compressed at the strain rate of 1.0~{$\mu$s$^{-1}$} (red squares), 10.0~{$\mu$s$^{-1}$} (blue triangles), and 100.0~{$\mu$s$^{-1}$} (black circles). Lines are the fitting to function $y=a+bx^{-2}$, where the coefficient $b$ increases with increasing strain rate. In particular, the coefficient of $b=40.6$~{\AA$^{2}$} for $\dot{\epsilon}=1.0$~{$\mu$s$^{-1}$} is only 6.7\% smaller than the prediction from the Euler buckling theory of $b=4\pi^{2}D/C_{11}=43.52$~{\AA$^{2}$}, where $D$ is the bending modulus and $C_{11}$ is the in-plane stiffness for SLMoS$_{2}$.}
  \label{fig_length}
\end{figure}

Fig.~\ref{fig_length} shows the length dependence of the critical strain in SLMoS$_{2}$ with width 40~{\AA}. This set of simulations are performed at a low temperature of 1.0~K, so that it can be consistent with the static Euler buckling theory. The system is compressed at the strain rate of 1.0~{$\mu$s$^{-1}$} (red squares), 10.0~{$\mu$s$^{-1}$} (blue triangles), and 100.0~{$\mu$s$^{-1}$} (black circles). According to the Euler buckling theory, the critical strain is inversely proportional to the square of the length of the SLMoS$_{2}$, so we fit simulation data to function $y=a+bx^{-2}$. The coefficient $b$ increases with increasing strain rate. In particular, the coefficient of $b=40.6$~{\AA$^{2}$} for $\dot{\epsilon}=1.0$~{$\mu$s$^{-1}$} agrees quite well with (error 6.7\%) the prediction from the Euler buckling theory of $b=4\pi^{2}D/C_{11}=43.52$~{\AA$^{2}$}, where $D$ is the bending modulus and $C_{11}$ is the in-plane stiffness for SLMoS$_{2}$. Furthermore, for the strain rate $\dot{\epsilon}=1.0$~{$\mu$s$^{-1}$}, the other coefficient $a$ is pretty small and there is almost no fluctuation between simulation data, both of which validate the usage of the small strain rate, $\dot{\epsilon}=1.0$~{$\mu$s$^{-1}$}. The critical strain for the shortest SLMoS$_{2}$ ($L=60$~{\AA}) deviates from the fitting curve in all of the three situations. It is due to the linear nature of the Euler buckling theory, while the nonlinear effect becomes important in the short system. \jwj{The critical strain does not depend on the width of the SLMoS$_{2}$, because the shape of the buckling mode is uniform in the width direction. Hence, the strain energy is the same in SLMoS$_{2}$ of different widths. That is why the width parameter does not present in the Euler buckling formula Eq.~(\ref{eq_strain2}).}

\begin{figure}[htpb]
  \begin{center}
    \scalebox{1}[1]{\includegraphics[width=8cm]{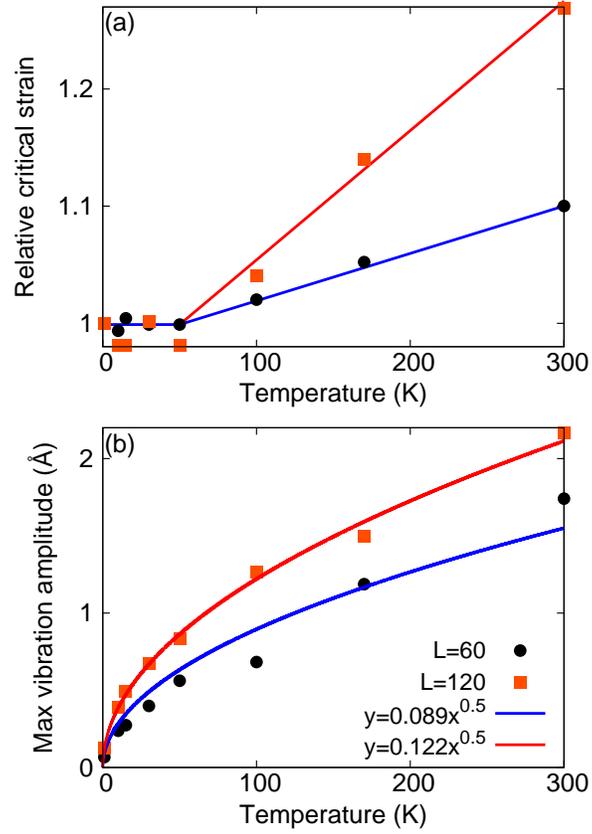}}
  \end{center}
  \caption{(Color online) The temperature effect on the critical strain for SLMoS$_{2}$ with width 40~{\AA} at strain rate 1.0~{$\mu$s$^{-1}$}. (a) Relative critical strain versus temperature for SLMoS$_{2}$ with $L=60$ and 120~{\AA}. The relative critical strain is scaled by the value at 1.0~K; i.e. 0.00939 and 0.00252 for $L=60$ and 120~{\AA} respectively. Lines are guided to the eye. (b) The maximum thermal vibration amplitude versus temperature for SLMoS$_{2}$ with $L=60$ (black circles) and 120~{\AA} (red squares). Lines are fitting to function $y=ax^{0.5}$.}
  \label{fig_temperature}
\end{figure}

Fig.~\ref{fig_temperature} shows the temperature effect on the critical strain for SLMoS$_{2}$ with width 40~{\AA} at strain rate 1.0~{$\mu$s$^{-1}$}. Panel (a) is the temperature dependence of the relative critical strain versus for SLMoS$_{2}$ with $L=60$ and 120~{\AA}. The relative critical strain is scaled by the value at 1.0~K; i.e. 0.00939 and 0.00252 for $L=60$ and 120~{\AA} respectively. Lines are guided to the eye. The critical strain in both systems keeps almost a constant at temperatures bellow 50~K. It increases linearly with increasing temperature for temperatures above 50~K, and the critical strain increases faster in longer SLMoS$_{2}$. This is due to strong thermal vibration at high temperatures. Panel (b) shows the maximum thermal vibration amplitude versus temperature for SLMoS$_{2}$ with $L=60$ (black circles) and 120~{\AA} (red squares). According to the equipartition theorem, the maximum thermal vibration amplitude can be fitted to function $y=ax^{0.5}$. Lines are the fitting results. The maximum vibration amplitude increases with increasing temperature. At higher temperatures (above 50~K), the maximum vibration amplitude is so large that it can disturb the buckling mode; i.e. the strong thermal vibration is able to heal the buckling-induced deformation in the SLMoS$_{2}$. At the initial buckling stage, atoms are displaced from its original position (following the buckling mode, i.e first lowest-frequency normal mode), but atoms are involved in a strong thermal vibration in the mean time. This large thermal vibration amplitude blurs the buckling deformation and the original configuration of the SLMoS$_{2}$. In other words, the thermal vibration introduces a healing mechanism for the buckling deformation at the initial buckling stage. Similar thermal healing mechanism has also been observed in the thermal treatment for defected carbon materials.\cite{KimYA2004cpl,JiangJW2010repair} Hence, larger compression strain is in need to intrigue the buckling phenomenon at higher temperatures.

\jwj{We end by noting that the present atomistic simulation actually has practical impact, although we emphasized the importance of the strain rate and temperature effects on the buckling of the SLMoS$_{2}$, which are technique aspects. MoS$_{2}$ and graphene have complementary physical properties, so it is natural to investigate the possibility of combining graphene and MoS$_{2}$ in specific ways to create heterostructures that mitigate the negative properties of each individual constituent. For example, graphene/MoS$_{2}$/graphene heterostructures have better photon absorption and electron-hole creation properties, because of the enhanced light-matter interactions by the single-layer MoS$_{2}$.\cite{BritnellL2013sci} It has been shown that the buckling critical strain for a graphene of 19.7~{\AA} in length is around 0.0068.\cite{LuQ2009ijam} From Euler buckling theorem, it can be extracted that the buckling critical strain for a graphene of 60.0~{\AA} in length is around 0.00073. Our simulations have shown that the buckling critical strain for SLMoS$_{2}$ of the same length is 0.0094, which is one order higher than graphene. It indicates that the SLMoS$_{2}$ can sustain stronger compression than graphene. The higher buckling critical strain for the SLMoS$_{2}$ is helpful for the graphene/MoS$_{2}$ heterostructure to protect from buckling damage.}

\section{Conclusion}
In conclusion, we have performed MD simulations to investigate the buckling of the SLMoS$_{2}$ under uniaxial compression. In particular, we examine the importance of the strain rate and temperature effects on the critical buckling strain. The critical strain increases linearly with increasing strain rate, and it keeps almost a constant at low temperatures. The critical strain increases with increasing temperature at temperatures above 50~K, because the strong thermal vibration is able to repair the buckling-induced deformation. The length dependence for the critical strain is in good agreement with the Euler buckling theory.

\textbf{Acknowledgements} The work is supported by the Recruitment Program of Global Youth Experts of China and the start-up funding from the Shanghai University.

%\bibliographystyle{aipnum4-1}
%\bibliography{/home/JiangJinWu/Documents/papers/mypapers/latex/biball}

\end{document}